# The $\pi\pi$ Amplitude in Generalized Chiral Perturbation Theory

**M. Knecht B. Moussallam** and **J. Stern**

*Division de Physique Théorique* [1], *Institut de Physique Nucléaire*
*F-91406 Orsay Cedex, France*

**Abstract**

The $\pi\pi$ interaction is studied at order one loop in the framework of Generalized Chiral Perturbation Theory.

*Contribution to the second edition of the DaΦne Physics Handbook, L. Maiani, G. Pancheri and N. Paver Eds., INFN, Frascati, to appear.*

[1] Unité de Recherche des Universités Paris XI et Paris VI associée au CNRS.

The general requirements of analyticity, unitarity and crossing symmetry, together with the Goldstone nature of the pion and isospin invariance, determine a low energy representation of the $\pi\pi$ scattering amplitude up to corrections of order $O([p/\Lambda_H]^8)$ [1], where $\Lambda_H$ denotes the scale at which particles other than pions can appear as intermediate states. This general representation depends on six independent low energy subtraction constants, which can be determined from experiment, given sufficiently precise data. Furthermore, only four of these constants (denoted as $\alpha$, $\beta$, $\lambda_1$ and $\lambda_2$ in the sequel) appear if two loop contributions are neglected [1]:

$$\begin{aligned} A(s|t,u) &= \frac{\beta}{F_\pi^2}\left(s-\frac{4}{3}M_\pi^2\right)+\alpha\frac{M_\pi^2}{3F_\pi^2} \\ &+ \frac{1}{F_\pi^4}(s-2M_\pi^2)^2\lambda_1 \\ &+ \frac{1}{F_\pi^4}\left[(t-2M_\pi^2)^2+(u-2M_\pi^2)^2\right]\lambda_2 \\ &+ \bar{J}_{(\alpha,\beta)}(s|t,u)+O(p^6/\Lambda_H^6)\,, \end{aligned} \tag{1}$$

with

$$\begin{aligned} \bar{J}_{(\alpha,\beta)}(s|t,u) &= \frac{1}{6F_\pi^4}\left\{4\left[\beta(s-\frac{4}{3}M_\pi^2)+\frac{5}{6}\alpha M_\pi^2\right]^2-\left[\beta(s-\frac{4}{3}M_\pi^2)-\frac{2}{3}\alpha M_\pi^2\right]^2\right\}\bar{J}(s) \\ &+ \frac{1}{12F_\pi^4}\left\{3\left[\beta(t-\frac{4}{3}M_\pi^2)-\frac{2}{3}\alpha M_\pi^2\right]^2+\beta^2(s-u)(t-4M_\pi^2)\right\}\bar{J}(t) \\ &+ \frac{1}{12F_\pi^4}\left\{3\left[\beta(u-\frac{4}{3}M_\pi^2)-\frac{2}{3}\alpha M_\pi^2\right]^2+\beta^2(s-t)(u-4M_\pi^2)\right\}\bar{J}(u)\,, \end{aligned} \tag{2}$$

and where $\bar{J}$ stands for the usual loop function,

$$16\pi^2\bar{J}(z)=\begin{cases} 2+\sigma(\ln\frac{1-\sigma}{1+\sigma}+i\pi) & ,\ \text{if } 4M_\pi^2\le z \\ 2-2\left(\frac{4M_\pi^2-z}{z}\right)^{\frac{1}{2}}\operatorname{arctg}(\frac{z}{4M_\pi^2-z})^{\frac{1}{2}} & ,\ \text{if } 0\le z\le 4M_\pi^2 \\ 2+\sigma\ln\frac{\sigma-1}{\sigma+1} & ,\ \text{if } z\le 0 \end{cases}\quad,\ \sigma(z)\equiv\sqrt{1-4M_\pi^2/z}\,. \tag{3}$$

The corresponding scattering lengths $a_l^I$ and slope parameters $b_l^I$ are given as follows:

$$\begin{aligned} a_0^0 &= \frac{1}{96\pi}\frac{M_\pi^2}{F_\pi^2}(5\alpha+16\beta)+\frac{5}{8\pi}\frac{M_\pi^4}{F_\pi^4}(\lambda_1+2\lambda_2)+\frac{1}{4608\pi^3}\frac{M_\pi^4}{F_\pi^4}(5\alpha+16\beta)^2 \\ a_0^2 &= \frac{1}{48\pi}\frac{M_\pi^2}{F_\pi^2}(\alpha-4\beta)+\frac{1}{4\pi}\frac{M_\pi^4}{F_\pi^4}(\lambda_1+2\lambda_2)+\frac{1}{1152\pi^3}\frac{M_\pi^4}{F_\pi^4}(\alpha-4\beta)^2 \\ a_1^1 &= \frac{1}{24\pi}\frac{1}{F_\pi^2}\beta-\frac{1}{6\pi}\frac{M_\pi^2}{F_\pi^4}(\lambda_1-\lambda_2)+\frac{1}{41472\pi^3}\frac{M_\pi^2}{F_\pi^4}(5\alpha^2-40\alpha\beta-16\beta^2) \\ a_2^0 &= \frac{1}{30\pi}\frac{1}{F_\pi^4}(\lambda_1+4\lambda_2)+\frac{1}{34560\pi^3}\frac{1}{F_\pi^4}(\alpha^2-48\beta^2) \\ a_2^2 &= \frac{1}{30\pi}\frac{1}{F_\pi^4}(\lambda_1+\lambda_2)+\frac{1}{7200\pi^3}\frac{1}{F_\pi^4}(\frac{1}{8}\alpha^2+\frac{13}{6}\alpha\beta-\frac{13}{3}\beta^2)\,, \end{aligned} \tag{4}$$

and

$$
\begin{aligned}
b_0^0 &= \frac{1}{4\pi}\frac{\beta}{F_\pi^2} + \frac{1}{\pi}\frac{M_\pi^2}{F_\pi^4}(2\lambda_1 + 3\lambda_2) + \frac{1}{3\pi^3}\frac{M_\pi^2}{F_\pi^4}(\frac{\beta^2}{3} + \frac{5}{96}\alpha\beta - \frac{5}{256}\alpha^2) \\
b_0^2 &= -\frac{1}{8\pi}\frac{\beta}{F_\pi^2} + \frac{1}{2\pi}\frac{M_\pi^2}{F_\pi^4}(\lambda_1 + 3\lambda_2) + \frac{1}{4608\pi^3}\frac{M_\pi^2}{F_\pi^4}(112\beta^2 - 8\alpha\beta - 7\alpha^2) \qquad (5) \\
b_1^1 &= -\frac{1}{6\pi}\frac{1}{F_\pi^4}(\lambda_1 - \lambda_2) + \frac{1}{4320\pi^3}\frac{1}{F_\pi^4}(\frac{47}{3}\beta^2 - \frac{65}{6}\alpha\beta - \frac{5}{24}\alpha^2) \ .
\end{aligned}
$$

The above parametrization of the low energy behaviour of the $\pi\pi$ amplitude and of the threshold parameters holds independently of more detailed considerations like, for instance, the value of the quark-antiquark condensate $B_0$ in the chiral limit. On the other hand, the expansions of $\alpha$, $\beta$, $\lambda_1$, $\lambda_2$ in powers of quark masses up to a given order will differ according to whether one assumes $B_0 \sim 1$ GeV, or $B_0 \sim 100$ MeV. In the sequel, we shall consider the amplitude $A(s|\,t,u)$ up to and including order $O(p^4)$ contributions only, within the framework of SU(3)×SU(3) Generalized Chiral Perturbation Theory (G$\chi$PT) [1, 2]. This means that both in Eq. (1) for the $\pi\pi$ scattering amplitude and in the expressions (4) and (5) for the threshold parameters, only the leading order expressions are to be retained for the parts that are quadratic in $\alpha$ and $\beta$. In particular, $\bar{J}_{(\alpha,\beta)}$ of Eq. (1) should be replaced by $\bar{J}_{(\alpha^{lead},\beta^{lead})}$ at that order, since the difference only affects contributions at orders $O(p^5)$ and $O(p^6)$. The leading order expressions $\alpha^{lead}$ and $\beta^{lead}$ and the higher order corrections are obtained from the effective lagrangian $\mathcal{L}^{eff} = \tilde{\mathcal{L}}^{(2)} + \tilde{\mathcal{L}}^{(3)} + \tilde{\mathcal{L}}^{(4)} + \cdots$ given in [2]. Expanding the contributions from $K\bar{K}$ and $\eta\eta$ intermediate states in powers of $s/M_P^2$, $P = K, \eta$ ($\Lambda_H < 2M_K,\ 2M_\eta$) and retaining only the dominant terms at order $O(p^4)$, one obtains the following result for $\lambda_1$ and for $\lambda_2$ ($\mu$ is an arbitrary subtraction scale):

$$
\begin{aligned}
\lambda_1 &= 4\,(2L_1^r(\mu) + L_3) - \frac{1}{48\pi^2}\left\{\ln\frac{M_\pi^2}{\mu^2} + \frac{1}{8}\ln\frac{M_K^2}{\mu^2} + \frac{35}{24}\right\} , \\
& \qquad\qquad\qquad\qquad\qquad\qquad\qquad\qquad\qquad\qquad (6) \\
\lambda_2 &= 4\,L_2^r(\mu) - \frac{1}{48\pi^2}\left\{\ln\frac{M_\pi^2}{\mu^2} + \frac{1}{8}\ln\frac{M_K^2}{\mu^2} + \frac{23}{24}\right\} .
\end{aligned}
$$

In these expressions, one recognizes the contributions coming from the tree diagrams, given in terms of the low energy constants $L_1$, $L_2$ and $L_3$, and the chiral logarithms coming from the loops. While these different contributions separately depend on the subtraction scale $\mu$, $\lambda_1$ and $\lambda_2$ are $\mu$-independent. The same holds for $\alpha$ and $\beta$, for which one obtains a similar decomposition:

$$
\alpha = \alpha_{tree}(\mu) + \alpha_{loop}(\mu) \ \ , \ \beta = \beta_{tree}(\mu) + \beta_{loop}(\mu) \ . \qquad (7)
$$

The tree level contributions read:

$$
\begin{aligned}
\alpha_{tree}(\mu) &= \alpha^{lead} \\
&+ \frac{1}{r-1}\left(\frac{F_K^2}{F_\pi^2} - 1\right)\left[2 + 12\zeta + 3\hat{\epsilon}(r-1) - 8\tilde{\xi}/\xi\right] \qquad (8) \\
&+ \delta\alpha_{tree}(\mu) \ ,
\end{aligned}
$$

with

$$\alpha^{lead} = 1 + 3\hat{\epsilon} \quad , \; \hat{\epsilon} \equiv 2\,\frac{r_2 - r}{r^2 - 1}(1 + 2\zeta) \quad , \; r_2 \; \equiv \; 2\,\frac{M_K^2}{M_\pi^2} - 1 \quad , \; \zeta \; \equiv \; Z_0^S/A_0 \; , \tag{9}$$

whereas

$$\beta_{tree} = \beta^{lead} + \frac{2}{r-1}\left(\frac{F_K^2}{F_\pi^2} - 1\right)\left(1 + 2\,\tilde{\xi}/\xi\right) + \delta\beta_{tree}(\mu) \quad , \; \beta^{lead} = 1 \; . \tag{10}$$

Notice that $\alpha^{lead}$ varies from 1 (the standard case) up to $\alpha^{lead} = 4$ for $r = r_1$, $r_1 \equiv 2M_K/M_\pi - 1$. In the above expressions, $\delta\alpha_{tree}(\mu)$ and $\delta\beta_{tree}(\mu)$ represent contributions from $\mathcal{L}_{(0,3)}$, $\mathcal{L}_{(0,4)}$ and $\mathcal{L}_{(2,2)}$. Their scale dependences are compensated by the $\mu$-dependences of the loop contributions, which read:

$$\begin{aligned}
\alpha_{loop}(\mu) \;\; = \;\; & -\frac{M_\pi^2}{32\pi^2 F_\pi^2}\left(\ln\frac{M_\pi^2}{\mu^2} + 1\right)\left[1 + 22\hat{\epsilon} + 33\hat{\epsilon}^2\right] \\
& -\frac{M_\pi^2}{32\pi^2 F_\pi^2}\left(\ln\frac{M_K^2}{\mu^2} + 1\right)\left[1 + \hat{\epsilon}(r+1)\right]\left[1 + 3\hat{\epsilon}(r+1)\right] \\
& -\frac{1}{3}\frac{M_\pi^2}{32\pi^2 F_\pi^2}\left(\ln\frac{M_\eta^2}{\mu^2} + 1\right)\left[1 + \hat{\epsilon}(2r+1) - \frac{2}{r-1}\cdot\Delta_{GMO}\right]^2 \\
& +\frac{\mu_\pi}{r-1}\left[1 - 18\zeta + \frac{3}{2}\hat{\epsilon}(5 - 13r - 16\zeta) + 20\tilde{\xi}/\xi\right] \\
& +\frac{\mu_K}{r-1}\left[2 + 12\zeta + 3\hat{\epsilon}(3+r) + 12\zeta\hat{\epsilon}(1+r) - 8\tilde{\xi}/\xi\right] \\
& +\frac{\mu_\eta}{r-1}\left[-3 + 6\zeta + \frac{1}{2}\hat{\epsilon}(r-1) + \frac{4}{r-1}(1+3\zeta)\cdot\Delta_{GMO} - 12\tilde{\xi}/\xi\right] \; ,
\end{aligned} \tag{11}$$

and

$$\begin{aligned}
\beta_{loop}(\mu) \;\; = \;\; & -\frac{1}{r-1}\left[5\mu_\pi - 2\mu_K - 3\mu_\eta\right]\left(1 + 2\tilde{\xi}/\xi\right) \\
& -\frac{M_\pi^2}{16\pi^2 F_\pi^2}\left(\ln\frac{M_\pi^2}{\mu^2} + 1\right)\left[2 + 5\hat{\epsilon}\right] \\
& -\frac{M_\pi^2}{32\pi^2 F_\pi^2}\left(\ln\frac{M_K^2}{\mu^2} + 1\right)\left[1 + \hat{\epsilon}(r+1)\right] \; ,
\end{aligned} \tag{12}$$

with $\Delta_{GMO}$ denoting the combination

$$\Delta_{GMO} = \left(3M_\eta^2 - 4M_K^2 + M_\pi^2\right)/M_\pi^2 \tag{13}$$

of pseudoscalar masses, and $\mu_P \equiv (32\pi^2 F_\pi^2)^{-1} M_P^2 \ln(M_P^2/\mu^2)$, $P = \pi, K, \eta$. In the standard case, $\delta\alpha_{tree}$ and $\delta\beta_{tree}$ are relegated to higher orders, while the quark mass ratio $r = m_s/\widehat{m}$ takes the value $r^{st}$, with [3]

$$r^{st} \equiv r_2 - 2\,\frac{M_K^2}{M_\pi^2}\,\Delta_M \ , \tag{14}$$

$$\Delta_M = -\mu_\pi + \mu_\eta + \frac{8}{F_\pi^2}\,(M_K^2 - M_\pi^2)\;(2L_8^r(\mu) - L_5^r(\mu)) \ . \tag{15}$$

To the order we consider here, the standard expressions for $\zeta$ and $\tilde{\xi}/\xi$ are

$$\zeta^{st} = \frac{L_6^r}{L_8^r}\ , \quad \left(\frac{\tilde{\xi}}{\xi}\right)^{st} = \frac{L_4^r}{L_5^r}\ , \tag{16}$$

while the splitting of the decay constants is expressed as [3]

$$\left(\frac{F_K^2}{F_\pi^2}\right)^{st} - 1 = 2\Delta_F \ , \tag{17}$$

with

$$\Delta_F = \frac{5}{4}\mu_\pi - \frac{1}{2}\mu_K - \frac{3}{4}\mu_\eta + \frac{4}{F_\pi^2}\,(M_K^2 - M_\pi^2) L_5^r \tag{18}$$

($\Delta_M$, $\Delta_F$ and $r^{st}$ are $\mu$-independent). The expressions (6) for $\lambda_1$ and for $\lambda_2$ remain unchanged, while one finds that $\alpha$ and $\beta$ become

$$\begin{aligned} \alpha^{st} &= 1 - 16\,\frac{M_\pi^2}{F_\pi^2}(2L_4^r + L_5^r) + 48\,\frac{M_\pi^2}{F_\pi^2}(2L_6^r + L_8^r) \\ &\quad - \frac{1}{32\pi^2}\,\frac{M_\pi^2}{F_\pi^2}\left\{\ln\frac{M_\pi^2}{\mu^2} + \ln\frac{M_K^2}{\mu^2} + \frac{1}{3}\ln\frac{M_\eta^2}{\mu^2} + \frac{7}{3}\right\} \end{aligned} \tag{19}$$

$$\begin{aligned} \beta^{st} &= 1 + 8\,\frac{M_\pi^2}{F_\pi^2}(2L_4^r + L_5^r) \\ &\quad - \frac{1}{32\pi^2}\,\frac{M_\pi^2}{F_\pi^2}\left\{4\ln\frac{M_\pi^2}{\mu^2} + \ln\frac{M_K^2}{\mu^2} + 5\right\} \ . \end{aligned} \tag{20}$$

These formulae could also be obtained directly from the standard expansion of the SU(3)×SU(3) effective lagrangian. The above exercise is a rather non trivial illustration of how the standard case arises as a special case of the generalized $\chi$PT. Notice also that restricting further the formulae (6), (19) and (20) to the SU(2)×SU(2) chiral limit reproduces the $O(p^4)$ $\pi\pi$ scattering amplitude of Ref. [3].

Let us now come back to the expressions of Eqs. (8) and (11) for $\alpha$ and Eqs. (10), (12) for $\beta$. Apart from the observable pseudoscalar masses and decay constants, they involve various other quantities which we discuss in turn:

**i)** As already mentioned earlier, $\delta\alpha_{tree}$ and $\delta\beta_{tree}$ collect tree level contributions from $\mathcal{L}_{(0,3)}$, $\mathcal{L}_{(0,4)}$ and $\mathcal{L}_{(2,2)}$. The various low energy constants involved in these pieces of the effective lagrangian, and which would appear only starting from order $O(p^6)$ in the standard case, are not under quantitative control at present. As an estimate of the uncertainties in $\alpha$ and in $\beta$ coming from our lack of knowledge of $\delta\alpha_{tree}$ and of $\delta\beta_{tree}$, we shall take the changes in $\alpha_{loop}(\mu)$ and in $\beta_{loop}(\mu)$ as the subtraction scale $\mu$ is varied between $\mu = M_\eta$ = 547.5 MeV and $\mu = M_\rho$ = 770 MeV.

**ii)** The two parameters $\zeta(= Z_0^S/A_0)$ and $\tilde{\xi}/\xi$ are also not known. Their values are expected to remain small as compared to unity, due to the Zweig rule. At leading order, the same vacuum stability argument that requires $B_0 \geq 0$ demands that $\zeta$ remains bounded as $r = m_s/\widehat{m}$ varies between $r_1$ and $r_2$,

$$0 \leq \zeta \leq \zeta_{crit}(r) \equiv \frac{1}{2}\frac{r-r_1}{r_2-r}\frac{r+r_1+2}{r+2} \ . \tag{21}$$

In what follows, we shall assume that these bounds still provide, at order $O(p^4)$, a good estimate of the uncertainty on $\zeta$, which we further restrict not to become larger than 0.5, i.e.,

$$0 \leq \zeta \leq \begin{cases} \zeta_{crit}(r) & , \text{ if } r \lesssim 14.5 \\ 0.5 & , \text{ if } r \geq 14.5 \end{cases} \ . \tag{22}$$

As for the remaining Zweig rule violating quantity $\tilde{\xi}/\xi$, we allow it to vary between −0.2 and +0.2,

$$\left|\frac{\tilde{\xi}}{\xi}\right| \lesssim 0.2 \ . \tag{23}$$

Notice that $\tilde{\xi}/\xi$ appears in the expression for the $\gamma\gamma \to \pi^0\pi^0$ cross section already at order one loop in G$\chi$PT [4], and also contributes to the $K_{l4}$ form factors, especially to $F$. Its value may in principle be obtained from very accurate $K_{e4}$ data, together with $L_1$, $L_2$ and $L_3$. Before discussing the values of these latter low energy constants, let us consider the ranges of values accessible to $\alpha$ and $\beta$ according to the preceeding discussion† . We show on Fig. 1 the bands of allowed values as functions of the quark mass ratio $r$. The bands delimited by the dotted curves correspond to the variations of $\zeta$ and of $\tilde{\xi}/\xi$ in the ranges (22) and (23), respectively, and to the estimates, via the $\mu$-dependences of $\alpha_{loop}$ and of $\beta_{loop}$, of the values and of the uncertainties associated to $\delta\alpha_{tree}(\mu)$ and to $\delta\beta_{tree}(\mu)$, respectively. The solid curves show the range of variation of $\alpha$ and of $\beta$ when only these last uncertainties are taken into account, with $\zeta$ and $\tilde{\xi}/\xi$ both taken to vanish. In the standard case, with the values of $L_4$, $L_5$, $L_6$ and $L_8$ as given in Ref. [3], we obtain, from Eqs. (19) and (20)

$$\alpha^{st} = 1.04 \pm 0.15 \quad , \qquad \beta^{st} = 1.08 \pm 0.03 \ , \tag{24}$$

†In all numerical evaluations, we take $M_\pi = M_{\pi^+}$ = 139.6 MeV, $M_K$ = 495.6 MeV, $M_\eta$ = 547.5 MeV, $F_\pi$ = 93.2 MeV and $F_K/F_\pi$ = 1.22.

in agreement with the values read from Fig. 1 for $r \sim 25$.

**iii)** The low energy constants $L_i$, $i$=1, 2, 3, can be extracted from the data on $K_{l4}$ decays [5, 6]. The expressions of the corresponding axial and vector current matrix elements are known at the one loop level both in the standard case [5, 7] and in generalized $\chi$PT [8]. In the latter case, the form factors depend on $r$ in addition. This dependence on the quark mass ratio may affect the values of the $L_i$'s one extracts from the data. In the standard case, the most recent analysis [7] of the data of Rosselet et al. [9] leads to the following values,

$$\begin{aligned} 10^3 L_1(M_\rho) &= 0.4 \pm 0.3 \\ 10^3 L_2(M_\rho) &= 1.35 \pm 0.3 \\ 10^3 L_3(M_\rho) &= -3.5 \pm 1.1 \ , \end{aligned} \tag{25}$$

which give

$$\lambda_1 = (-6.4 \pm 6.8) \times 10^{-3} \quad , \qquad \lambda_2 = (10.8 \pm 1.2) \times 10^{-3} \ . \tag{26}$$

A preliminary study of the generalized case shows that these values of $L_1$, $L_2$ and $L_3$ tend to decrease in absolute value with $r$ [8]. The constant $L_3$ is the most sensitive to variations in the quark mass ratio $r$ and its central value becomes $10^3 L_3(M_\rho) \sim -2$, for $r \lesssim 10$, with an uncertainty comparable to the one shown in Eq. (25). The variations in the values of $L_1$ and of $L_2$, however, are smaller and affected by larger error bars. Within these, they remain compatible with the standard values (25). For $L_3 = -2 \times 10^{-3}$ and for $L_1$, $L_2$ as before, $\lambda_2$ remains unchanged, but $\lambda_1$ decreases (in absolute value) by $\sim$30% and its central value becomes $\lambda_1 = -4.4 \times 10^{-3}$.

In Figs. 2 to 4 we have plotted the behaviours of the scattering lengths $a_0^0$ and $a_1^1$, and of the slope parameter $b_0^0$ (all in units of $M_\pi^+$) as the quark mass ratio $m_s/\widehat{m}$ varies between 8 and 30. The two error bands (solid lines and dotted lines) come from the corresponding errors on $\alpha$ and on $\beta$ shown in Fig. 1. The plots correspond to the central values of the $L_i$'s, $i = 1, 2, 3$, as given by Eq. (25) above. In Fig. 5 we show the difference $a_0^0 - a_0^2$, which can be obtained directly from the lifetime measurement of $\pi^+\pi^-$ atoms, an experiment planed at CERN [10].

The $\pi\pi$ phase shift combination $\delta_0^0 - \delta_1^1$ can be extracted down to very small energies by analyzing $K_{l4}$ decay data. In order to calculate the phase shifts, one first constructs the amplitude $f_l^I(s)$ for a given partial wave $l$ and isospin $I$ from $A(s|t,u)$ (see e.g. the appendix of Ref [1]). At order $O(p^4)$ the phase shifts are given as usual by

$$\delta_l^I(s) = \sqrt{1 - \frac{4M_\pi^2}{s}}\, \Re e\ f_l^I(s) + O(p^6/\Lambda_H^6) \ . \tag{27}$$

This expression for $\delta_0^0(s)$ should only be used close to the threshold, since the perturbative $O(p^4)$ amplitude $f_0^0(s)$ violates unitarity for energies $\sqrt{s}$ above 430-440 MeV. The predictions of the G$\chi$PT for $\delta_0^0 - \delta_1^1$ is shown in Fig. 6, for $r = 10$, and compared to the data of Rosselet et al. [9]. The predictions of the standard $\chi$PT are also plotted. In the latter case, we use the values (25) of the low energy constants $L_{1,2,3}$, and take for $\alpha^{st}$ and $\beta^{st}$ the central values in Eq. (24), whereas the leading order values of $\alpha^{st}$ and $\beta^{st}$ are both equal to 1. At present, both results are compatible with the data, but it is clear that a reduction of the experimental error bars by a factor of two could be enough to make the distinction between the alternatives $B_0 \sim \Lambda_H$ and $B_0 \sim F_0$. Notice also that the uncertainties materialized by the dotted lines in Fig. 6 include the variations of the values of $L_3$ between $-3.5 \times 10^{-3}$ and $-2 \times 10^{-3}$. The solid lines, which represent the variations induced by changing the subtraction scale $\mu$ in $\alpha_{loop}$ and in $\beta_{loop}$ between $M_\eta$ and $M_\rho$, correspond to $L_3 = -3.5 \times 10^{-3}$. For $L_3 = -2 \times 10^{-3}$, the solid lines would lie at the center of the dotted band.

Finally, we have gathered, in Table 1, the central values of the threshold parameters of Eqs. (4) and (5), for $r = 10$. The value $r = 10$ taken in the table below and in Fig. 6 showing the phase shifts, is suggested both by the analysis of the deviation from the Goldberger-Treiman relation [11], and by a recent analysis of the process $\gamma\gamma \to \pi^0\pi^0$ in G$\chi$PT [4].

| | experiment | standard $\chi$PT | generalized $\chi$PT for $r$=10 | |
|---|---|---|---|---|
| | [12] | [5] | $10^3 L_3 = -3.5$ | $10^3 L_3 = -2$ |
| $a_0^0$ | 0.26±0.05 | 0.20 | 0.27 | 0.28 |
| $b_0^0$ | 0.25±0.03 | 0.25 | 0.26 | 0.28 |
| $-10a_0^2$ | 0.28±0.12 | 0.41 | 0.23 | 0.21 |
| $-10b_0^2$ | 0.82±0.08 | 0.72 | 0.79 | 0.75 |
| $10a_1^1$ | 0.38±0.02 | 0.37 | 0.39 | 0.38 |
| $10^2 b_1^1$ | | 0.48 | 0.48 | 0.28 |
| $10^2 a_2^0$ | 0.17±0.03 | 0.18 | 0.18 | 0.21 |
| $10^3 a_2^2$ | 0.13±0.3 | 0.20 | 0.24 | 0.57 |

**Table 1:** *Results for the threshold parameters for $r = 10$ and for two values of $L_3$, as compared to the standard predictions. The values shown correspond to the central values $\alpha = 3$, $\beta = 1.19$, $\lambda_1 = -6.4 \times 10^{-3}$ (for $L_3 = -3.5 \times 10^{-3}$) or $\lambda_1 = -4.4 \times 10^{-3}$ (for $L_3 = -2 \times 10^{-3}$), and $\lambda_2 = 10.8 \times 10^{-3}$.*

The main differences between the standard predictions and the generalized case for $r = 10$ arise in $a_0^0$ (30%), in $a_0^2$ (50%) and in $b_1^1$ (for the lowest value of $|L_3|$).

In all the figures presented here, the error bands delimited by the dotted lines are in principle reducible, if our knowledge of the two Zweig rule violating parameters $\zeta$ and $\tilde{\xi}/\xi$ were to improve. A simultaneous determination of $r$ and of $\zeta$ is in principle possible from very accurate data on both $\pi\pi$ and $\pi K$ scattering. As for the value of $\tilde{\xi}/\xi$, it might be obtained from more precise data on $K_{e4}$ decays and on $\gamma\gamma \to \pi^0\pi^0$. The smaller error bands, delimited by the solid curves, reflect our lack of information concerning the low energy constants of $\mathcal{L}_{(0,3)}$, $\mathcal{L}_{(0,4)}$ and $\mathcal{L}_{(2,2)}$, and a significant improvement seems unlikely so far. Living with only these last uncertainties still offers a good possibility to desentangle the low $B_0$ and large $B_0$ alternatives, provided the experimental data become accurate enough. On the theoretical side, it is in principle possible to study the effect of higher orders on the values of the parameters $\alpha$, $\beta$ and $\lambda_i$, $i$=1, 2, by comparing a fit to the data obtained from the parametrization of the amplitude given in Eq. [1], with a similar fit done by using the general parametrization of $A(s|t, u)$ up to and including the order two loop given in Ref. [1].

# References

[1] J. Stern, H. Sazdjian and N. H. Fuchs, Phys. Rev. **D47** (1993) 3814.

[2] M. Knecht and J. Stern, Generalized Chiral Perturbation Theory, preprint IPNO/TH 94-53, contributed to the second edition of the DaΦne Physics Handbook, L. Maiani, G. Pancheri and N. Paver Eds., INFN, Frascati, to appear.

[3] J. Gasser and H. Leutwyler, Ann. Phys. **158** (1984) 142; Nucl. Phys. **B250** (1985) 465.

[4] M. Knecht, B. Moussallam and J. Stern, Nucl. Phys. **B429** (1994) 125.

[5] J. Bijnens, Nucl. Phys. **B337** (1990) 635.

[6] C. Riggenbach, J. Gasser, J.F. Donoghue and B.R. Holstein, Phys. Rev. **D43** (1991) 127.

[7] J. Bijnens, G. Colangelo and J. Gasser, $K_{l4}$-decays beyond one loop, preprint BUTP-94/4 and ROM2F 94/05 (hep-ph/9403390), Nucl. Phys. B, in press.

[8] M. Knecht and J. Stern, to appear, and contribution to the *EurodaΦne Collaboration Meeting*, April 19-22 1994, Frascati (Italy).

[9] L. Rosselet et al., Phys. Rev. **D15** (1977) 574.

[10] G. Czapek et al., Letter of intent: Lifetime measurements of $\pi^+\pi^-$ atoms to test low-energy QCD predictions, preprint CERN/SPSLC 02-44.

[11] N. H. Fuchs, H. Sazdjian and J. Stern, Phys. Lett. **B238** (1990) 380.

[12] M. M. Nagels et al., Nucl. Phys. **B147** (1979) 189.

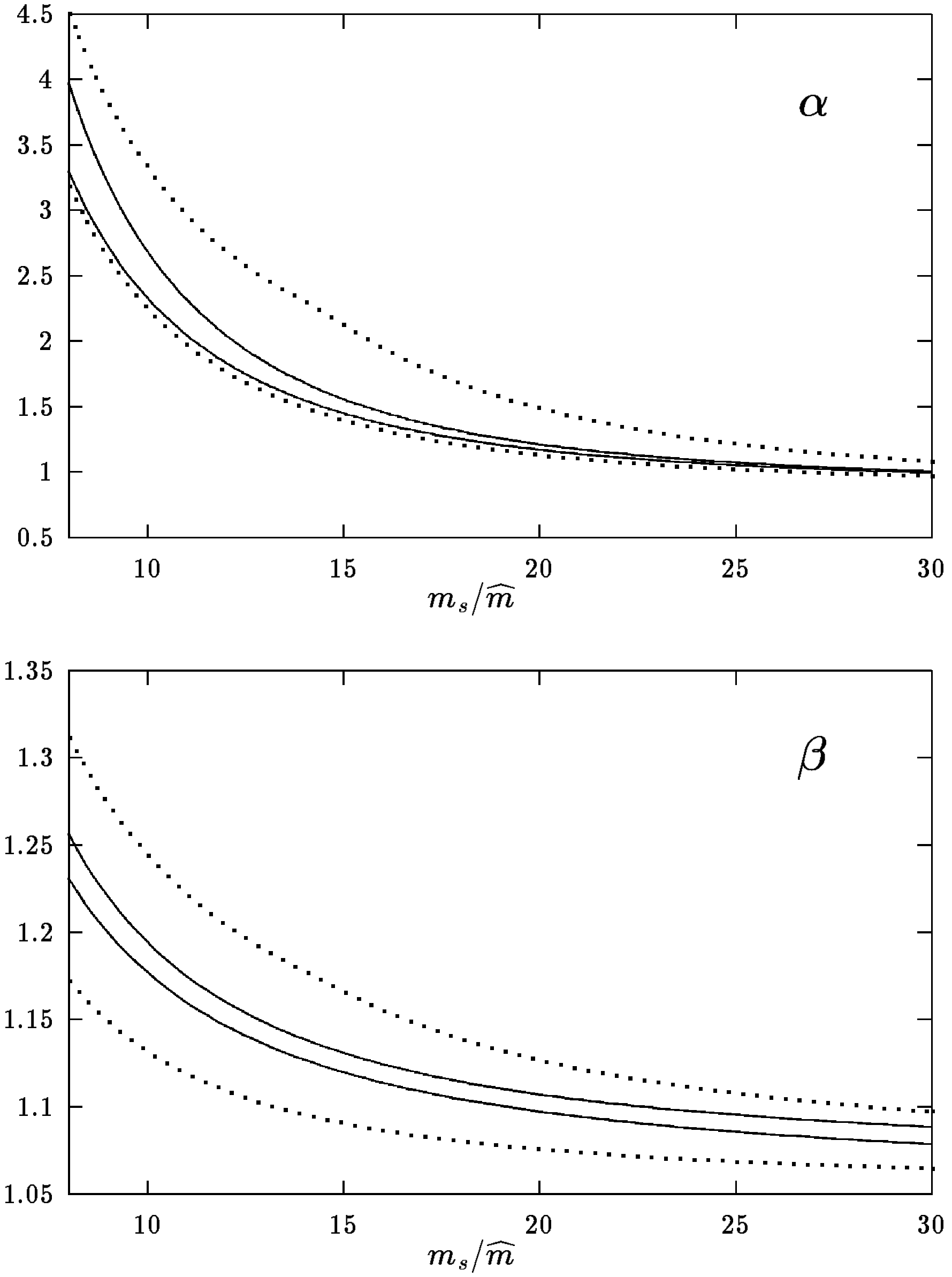

Figure 1: The bands of allowed values of $\alpha$ and of $\beta$ as functions of the quark mass ratio $m_s/\widehat{m}$. The dotted curves correspond to variations of $\zeta$ and of $\tilde{\xi}/\xi$ in the ranges set by Eqs. (22) and (23), respectively, and with $\delta\alpha_{tree}(\mu)$ and $\delta\beta_{tree}(\mu)$ estimated by the respective variations of $\alpha_{loop}(\mu)$ and of $\beta_{loop}(\mu)$ for $\mu$ between 547.5 MeV and 770 MeV. The solid curves show the allowed values of $\alpha$ and of $\beta$ when only the latter uncertainties are taken into account, having put $\zeta$ and $\tilde{\xi}/\xi$ to zero.

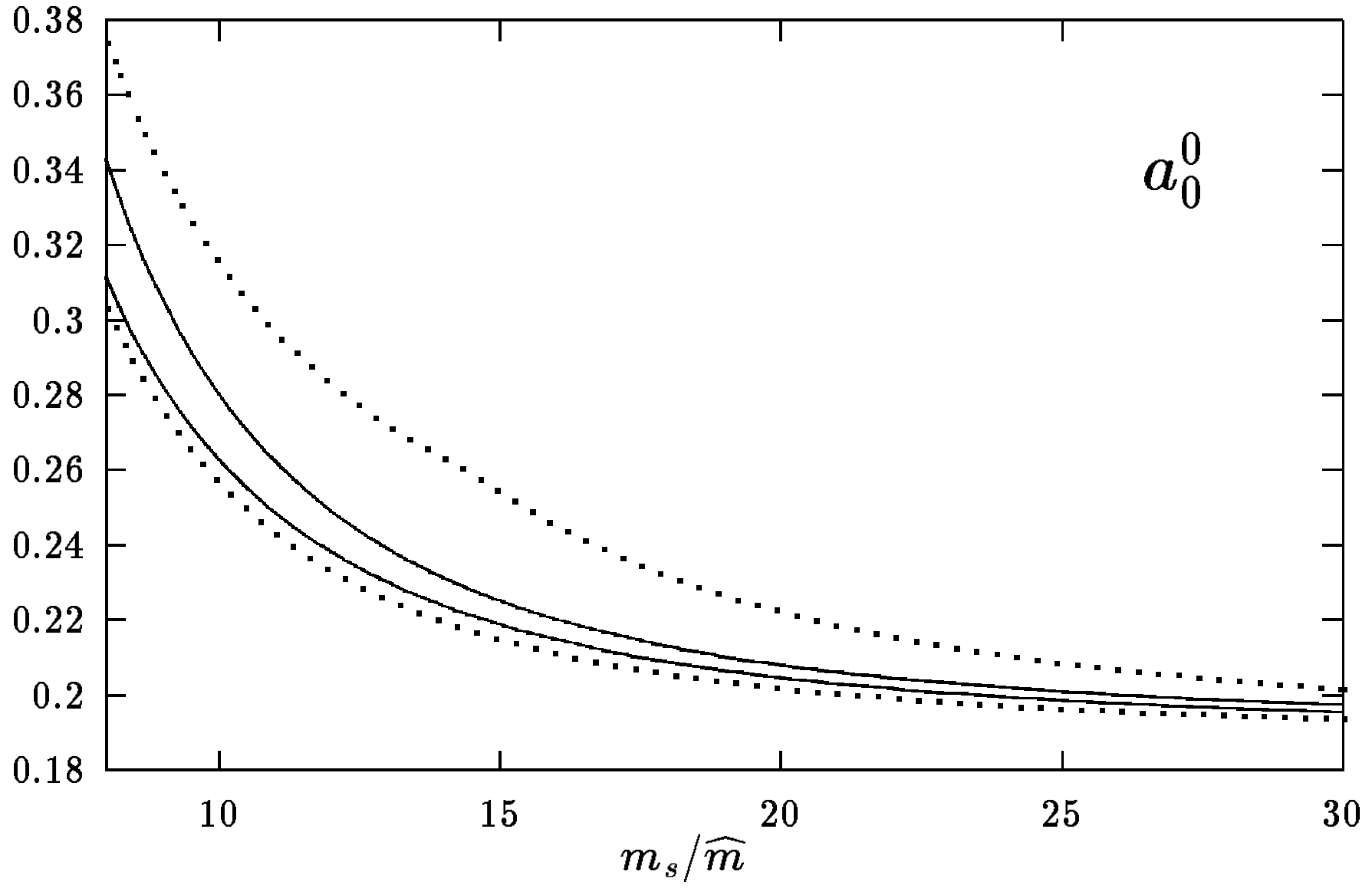

Figure 2: The I=0 S-wave scattering lenth $a_0^0$ as $r$ varies between 8 and 30. The band delimited by the solid lines and the band delimited by the dotted lines arise from the corresponding uncertainties in $\alpha$ and in $\beta$ shown in Fig. 1.

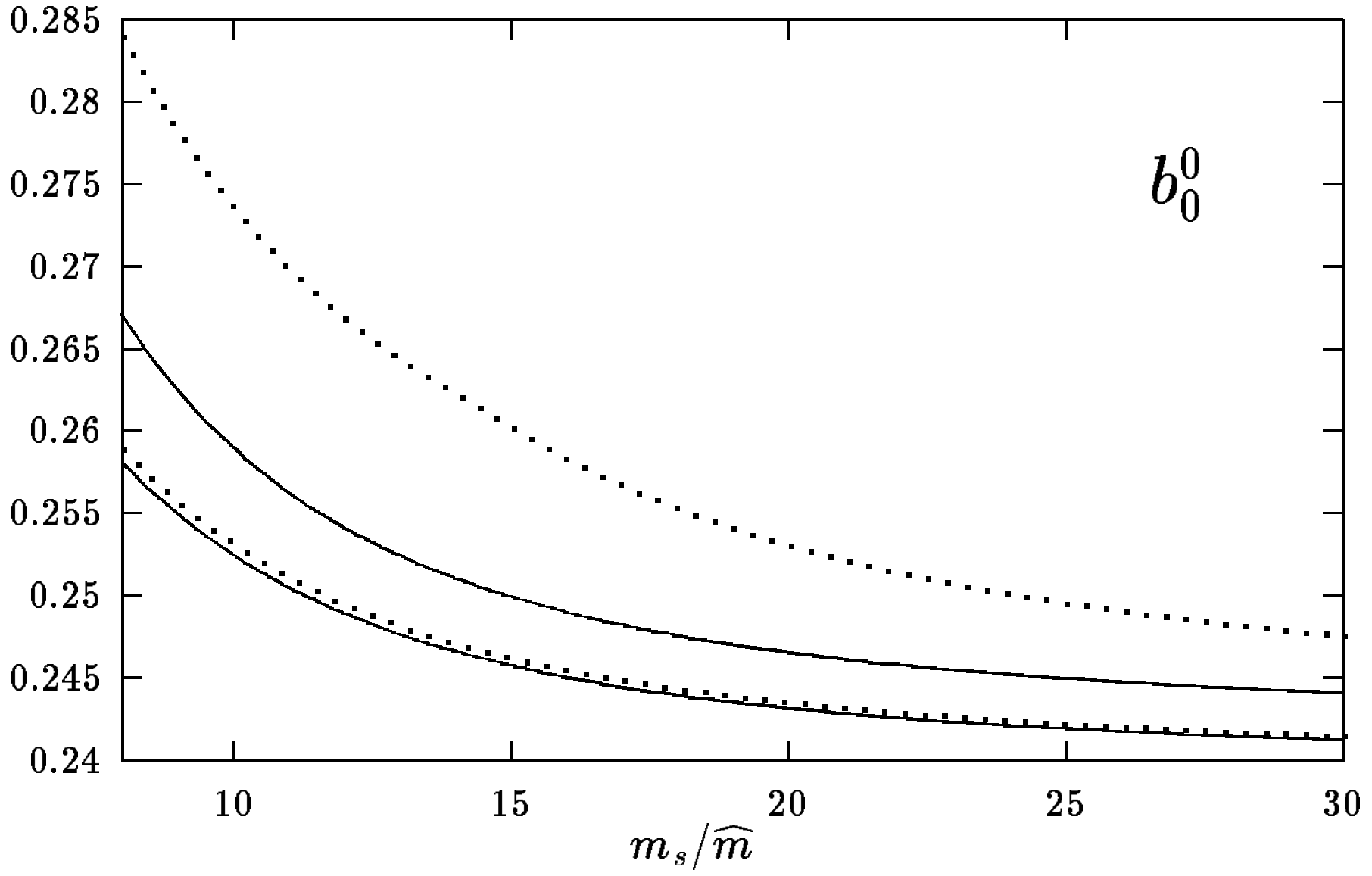

Figure 3: The I=0 S-wave slope parameter $b_0^0$ as $r$ varies between 8 and 30. The band delimited by the solid lines and the band delimited by the dotted lines arise from the corresponding uncertainties in $\alpha$ and in $\beta$ shown in Fig. 1.

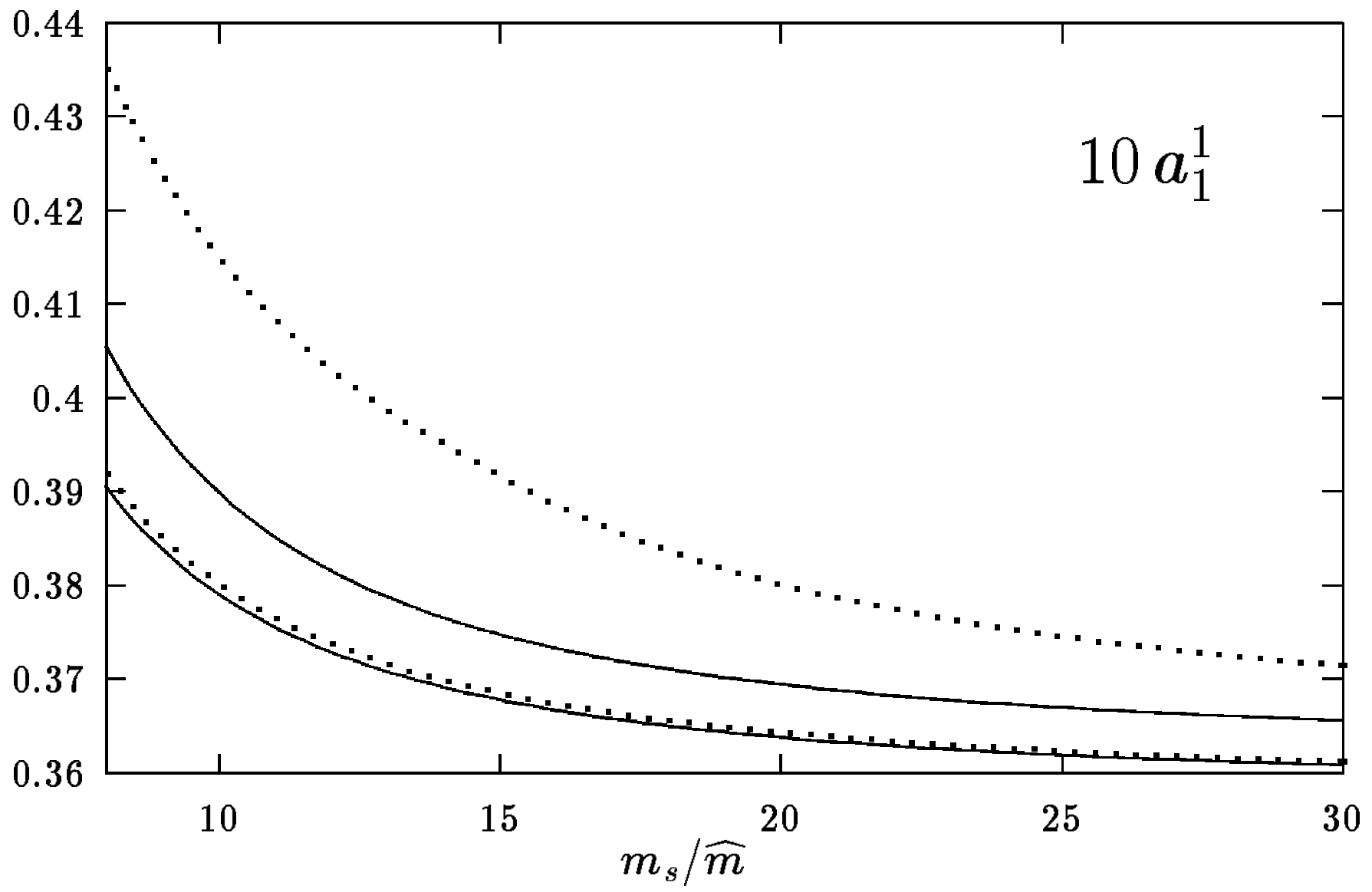

Figure 4: The I=1 P-wave scattering lenth $a_1^1$ as $r$ varies between 8 and 30. The band delimited by the solid lines and the band delimited by the dotted lines arise from the corresponding uncertainties in $\alpha$ and in $\beta$ shown in Fig. 1.

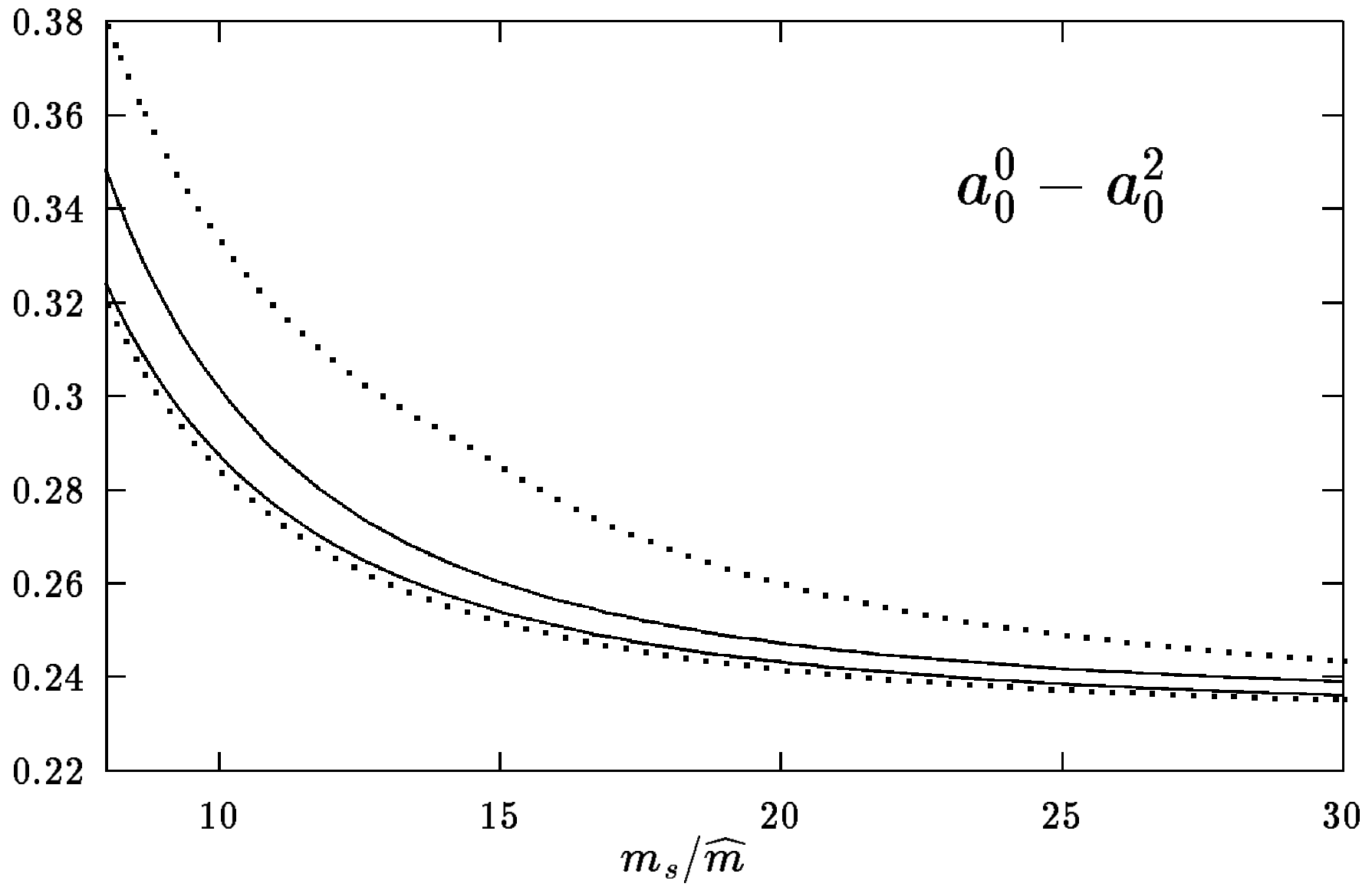

Figure 5: The difference $a_0^0 - a_0^2$ of S-wave scattering lengths as a function of $r$.

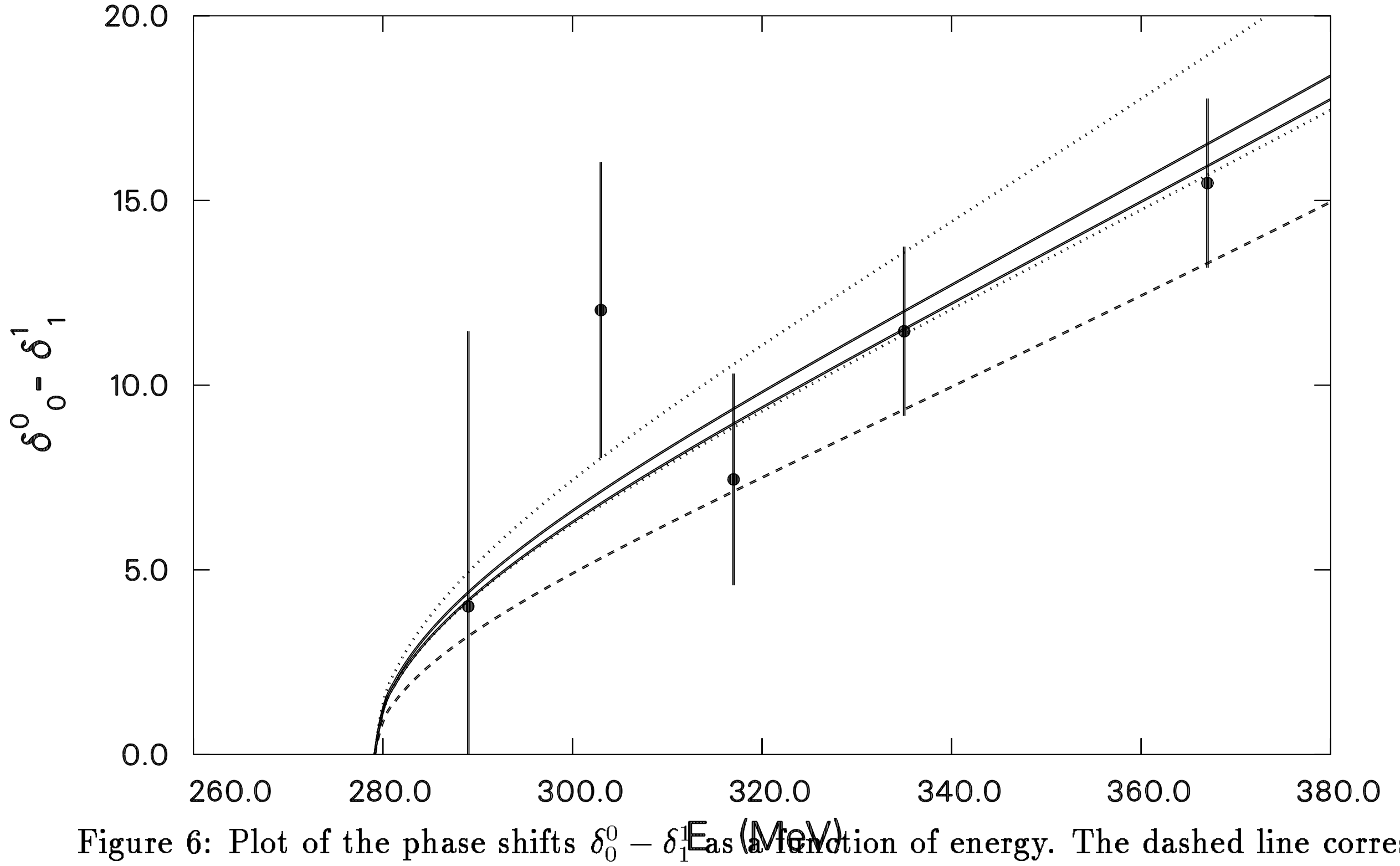

Figure 6: Plot of the phase shifts $\delta_0^0 - \delta_1^1$ as a function of energy. The dashed line corresponds to the predictions of the standard case. The solid and dotted lines give the allowed values for $r = 10$ in the generalized case (see text). The data points are from Rosselet et al. [8].